\documentclass[letterpaper]{article} 
\usepackage{aaai2026}  
\usepackage{times}  
\usepackage{helvet}  
\usepackage{courier}  
\usepackage[hyphens]{url}  
\usepackage{graphicx} 
\usepackage{natbib}  
\usepackage{caption} 
\usepackage{tikz}
\usepackage{pgfplots}
\usepackage{amssymb}
\usepackage{amsmath}
\usepackage{amsthm}
\usepackage{algpseudocode}
\usepackage{algorithm}
\usepackage{caption}
\usepackage{subcaption}
\usepackage{enumitem}
\usepackage{booktabs}

\newcommand{\semG}[2]{\bigl[\!\![#1]\!\bigr]^{G}_{#2}}

\usepackage{newfloat}
\usepackage{listings}

\newtheorem{definition}{Definition}
\newtheorem{proposition}{Proposition}
\newtheorem{example}{Example}
\newtheorem{theorem}{Theorem}

\newif\ifdraft\drafttrue

\newcommand{\init}{{I}}

\newcommand{\last}{\mbox{last}}

\newcommand{\APf}{\textnormal{Ap}}
\renewcommand{\phi}{\varphi}

\newcommand{\Prop}{\APf} 
\newcommand{\until}{{\bf U}}
\newcommand{\U}{\until\xspace}
\newcommand{\release}{{\bf R}}
\newcommand{\R}{\release}

\newcommand{\X}{{\bf X}}
\newcommand{\G}{{\bf G}}

\newcommand{\coop}[1]{\langle\!\langle{#1}\rangle\!\rangle}

\newcommand{\coalition}{A}

\newcommand{\Ag}{\textnormal{Ag}}
\newcommand{\ag}{a}

\newcommand{\Act}{\textnormal{Act}}

\newcommand{\act}{c}

\newcommand{\setpos}{St}

\newcommand{\pos}{st}
\newcommand{\CGS}{\relax\ifmmode\mathcal G\else$\textrm{CGS}$\xspace\fi}

\newcommand{\transCGS}{t}
\newcommand{\availability}{d}
\newcommand{\val}{\ell}

\newcommand{\out}{\textnormal{out}}

\newcommand{\Agt}{\mathbb{A}\textnormal{gt}}

\newcommand{\wCGS}{\relax\ifmmode \mathcal G\else$\textrm{wCGS}$\xspace\fi}
\newcommand{\rfCGS}{\relax\ifmmode \mathcal G\else$\textrm{rfCGS}$\xspace\fi}
\newcommand{\wCGT}{\relax\ifmmode \mathcal T\else$\textrm{wCGT}$\xspace\fi}

\renewcommand{\Gamma}{A}
\renewcommand{\sigma}{s}

\protected\def\tpath{\ifmmode \spath \else path\xspace\fi}

\newcommand{\struct}{\mathcal S}
\newcommand{\KS}{\relax\ifmmode \struct\else\textrm{wKS}\xspace\fi}

\newcommand{\spath}{\lambda}

\newcommand{\Bool}{\text{Bool}}

\newcommand{\tauto}{\mathcal A}
\newcommand{\Nauto}{\ifmmode \mathcal N \else NPT\xspace\fi}
\newcommand{\Uauto}{\ifmmode \mathcal U \else UPT\xspace\fi}

\newcommand{\auto}{\ifmmode \tauto \else APT\xspace\fi}

\newcommand\LTL{\textnormal{\sffamily LTL}\xspace}

\newcommand\CTL{\textnormal{{\sffamily CTL}}\xspace}

\newcommand\ATL{\textnormal{{\sffamily ATL}}\xspace}

\newcommand\NatATLr{\textnormal{{\sffamily NatATL}}_{\textnormal{r}}\xspace}

\newcommand\ATLr{\textnormal{{\sffamily ATL}}_{\textnormal{r}}\xspace}
\newcommand\ATLR{\textnormal{{\sffamily ATL}}_{\textnormal{R}}\xspace}

\newcommand\SL{\textnormal{{\sffamily SL}}\xspace}
\newcommand{\Fu}[2]{\ensuremath {{#1}{[#2]}}\xspace}
\newcommand{\Func}{\ensuremath{\mathcal F}}

\newcommand{\SLF}{\Fu{\SL}{\Func}}
\newcommand{\ATLF}{\Fu{\ATL}{\Func}}

\newcommand{\CTLF}{\Fu{\CTL}{\Func}}

\newcommand{\HumanATLF}{%
  \textnormal{\sffamily Human}\Fu{\ATL}{\Func}%
}

\newcommand{\HumanATLrF}{%
  \textnormal{\sffamily Human}\Fu{\ATLr}{\Func}%
}

\newcommand{\HumanATLRF}{%
  \textnormal{\sffamily Human}\Fu{\ATLR}{\Func}%
}

\newcommand{\profile}[1]{\boldsymbol{#1}}

\newcommand{\lowb}{0}

\protected\def\PTIME{\ifmmode \mbox{\sc P} \else {\sc P}\xspace\fi}
\protected\def\PSPACE{\ifmmode \mbox{\sc Pspace} \else {\sc Pspace}\xspace\fi}
\protected\def\EXPTIME{\ifmmode \mbox{\sc Exptime} \else {\sc Exptime}\xspace\fi}
\protected\def\NONELEMENTARY{\ifmmode \mbox{\sc NonElementary} \else {\sc NonElementary}\xspace\fi}
\protected\def\ELEMENTARY{\ifmmode \mbox{\sc Elementary} \else {\sc Elementary}\xspace\fi}

\renewcommand{\ldots}{...}
\renewcommand{\dots}{...}

\algnewcommand\algorithmicguess{\textbf{Guess}}
\algnewcommand\algorithmicendguess{\textbf{EndGuess}}
\algblock{Guess}{EndGuess}

\usepackage{newfloat}    
\usepackage{caption}     
\usepackage{xspace}

\usepackage{stmaryrd}

\DeclareFloatingEnvironment[
  fileext=lst,
  listname={Listings},
  name=Listing,
  placement=tb
]{listing}

\DeclareCaptionStyle{ruled}{labelfont=normalfont,labelsep=colon,strut=off}
\title{When Natural Strategies Meet Fuzziness and Resource-Bounded Actions

(Extended Version)}
\author {
    Marco Aruta\textsuperscript{\rm 1},
    Francesco Improta\textsuperscript{\rm 1},
    Vadim Malvone\textsuperscript{\rm 2},
    Aniello Murano\textsuperscript{\rm 1}
}
\affiliations {
    \textsuperscript{\rm 1}University of Naples Federico II, Naples, Italy\\
    \textsuperscript{\rm 2}LTCI, Télécom Paris, Institut Polytechnique de Paris, Palaiseau, France\\
     marco.aruta@unina.it, francesco.improta@unina.it, aniello.murano@unina.it, vadim.malvone@telecom-paris.fr
}

\usepackage{bibentry}

\begin{document}

\maketitle

\begin{abstract}
In formal strategic reasoning for Multi-Agent Systems (MAS), agents are typically assumed to (i) employ arbitrarily complex strategies, (ii) execute each move at zero cost, and (iii) operate over fully crisp game structures. These idealized assumptions stand in stark contrast with human decision-making in real-world environments. The \emph{natural strategies} framework along with some of its recent variants, partially addresses this gap by restricting strategies to concise rules guarded by regular expressions. Yet, it still overlook both the cost of each action and the uncertainty that often characterizes human perception of facts over the time. In this work, we introduce \HumanATLF, a logic that builds upon natural strategies employing both fuzzy semantics and resource‐bound actions: each action carries a real-valued cost drawn from a non‐refillable budget, and atomic conditions and goals have degrees in [0,1]. We give a formal syntax and semantics, and prove that model checking is in P when both the strategy complexity $k$ and resource budget $b$ are fixed, NP-complete if just one strategic operator over Boolean objectives is allowed, and $\Delta^P_2$‐complete when $k$ and $b$ vary. Moreover, we show that recall‐based strategies can be decided in PSPACE. We implement our algorithms in VITAMIN, an open source model-checking tool for MAS and validate them on an adversarial resource-aware drone rescue scenario.
\end{abstract}

\begin{links}
    \link{Code}{https://github.com/MarcoAruta/HumanATLFTool}
\end{links}



\section{\uppercase{Introduction}}
\label{sec:introduction}

Formal strategic reasoning in Multi-Agent Systems (MAS) plays a central role in verifying and synthesizing behaviors in distributed and adversarial settings. A milestone in this area is \emph{Alternating-time Temporal Logic} (\ATL), introduced by Alur, Henzinger, and Kupferman \cite{alur2002alternating}, which provides a formal framework to express and verify the strategic abilities of coalitions of agents. \ATL and its extensions have greatly advanced the formal aspects of strategic reasoning, enabling model checking techniques that scale to complex MAS \cite{van2005logic,aagotnes2006action,aagotnes2007alternating, schnoor2009probabilistic,della2018parity, DBLP:journals/logcom/MogaveroMV16, knapik2019timed, jamroga2025probabilistic}. Over the years, theoretical developments and practical implementations have further refined these foundations, improving expressiveness, decidability, and verification algorithms that bridge logic, game theory, and automated reasoning \cite{chatterjee2010strategy,mogavero2013boundary, DBLP:journals/tocl/MogaveroMPV14,mogavero2017reasoning, aminof2018graded,belardinelli2022reasoning}. 

Although \ATL comes with a powerful strategic-reasoning capability for MAS, it assumes idealized agents with unlimited reasoning capacity, making it unsuitable for modeling the bounded rationality typical of human decision-making. This has spurred the community to look for a new strategic reasoning framework that better represents (and allows to reproduce) bounded, human-like reasoning. This has led to \emph{natural strategies} \cite{JMM19a,DBLP:conf/ijcai/KaminskiKJ25,belardinelli2019strategy,jamroga2022measure}, which allows to define human-reproducible strategies represented by means of rules guarded by regular expressions.

While natural strategies allow to capture the important aspect of bounded rationality, it  measures only the combinatorial cost of evaluating propositional conditions; it does not take into account that in real‐world strategic reasoning settings, humans (and in general resource‐limited AI agents) cannot maintain arbitrarily complex strategies. In realistic scenarios, each strategy action consumes tangible resources, whether cognitive effort or physical energy. As an example, consider a fleet of delivery drones operating in a smart city: every “ascend” or “move” maneuver depletes a fixed portion of the battery. A path that seems admissible under Boolean or regular‐expression guards may prove infeasible if the sum of individual action costs exceeds the drone’s remaining charge. At the same time, decision makers frequently employ graded judgments rather than strict true–false distinctions: goals such as “maintain safe distance” or “minimize detection risk” are naturally expressed in terms of satisfaction levels. By endowing a strategic logic with \emph{fuzzy} semantics (i.e. the truth degrees in $[0,1]$) one can capture these intermediate satisfaction levels and reason about partial achievements of objectives. We contend that a unified treatment of human‐feasible strategy complexity, action‐level resource consumption, and fuzzy outcome evaluation delivers a more faithful model of strategic reasoning in human-like agents.

\paragraph{Our Contribution.}  
We introduce \HumanATLF, a Human-like logic that extends \ATL on three key quantitative dimensions. First, strategy descriptions are human‐readable natural strategies: each rule is a simple guarded‐action pair whose syntactic complexity is capped by a parameter $k$. Second, every action draws from a finite, non‐refillable budget $b$, ensuring plans respect explicit resource constraints. Third, objectives are evaluated under fuzzy semantics, assigning truth degrees in $[0,1]$ to capture partial fulfillment of the aforementioned goals. On this foundation, we develop the full syntax and fuzzy, cost‐aware semantics of \HumanATLF\ and then analyze the computational complexity of the model-checking problem. The latter established three key results: the verification is in P when both the strategy complexity bound $k$ and the available budget $b$ are fixed; following that, even a single strategic operator over Boolean objectives induces NP-completeness; and finally, that model checking becomes $\Delta^P_2$-complete when $k$ and $b$ are treated as part of the input. In the case of recall-based strategies, we show that model checking remains tractable by exploring a bounded unfolding of the game graph that respects the fixed complexity bound, demonstrating that it is in PSPACE. To demonstrate practicality, we implement our algorithms within the open-source VITAMIN tool \cite{DBLP:journals/corr/abs-2403-02170} and evaluate their performance on a suite of benchmarks, including an adversarial drone-rescue scenario designed to highlight the interplay between fuzzy satisfaction, energy limitations, and natural strategy construction. This comprehensive framework aims to offer the first end-to-end approach to strategic verification that unifies interpretability, quantitative reasoning, and cost-awareness within a single, executable formalism.

\paragraph{Related Works.}
The seminal quantitative Alternating‐Time Temporal Logic first provided by Jamroga in 2008 \cite{jamroga2008temporal} helped to blossom interest in the interaction between logics for strategic reasoning and fuzziness. The other foundational work that inspired our approach is the contribution of Alechina et al. \cite{10.5555/1838206.1838274} where they reason about the availability of resources and the lack of a straightforward way of reasoning about resource requirements in \ATL \cite{alur2002alternating}. Subsequent works explored the impact of restrictions such as memory bounds and imperfect information on expressiveness and decidability \cite{DBLP:journals/tocl/MogaveroMPV14,DBLP:conf/tacas/ChenFKPS13,DBLP:conf/aaai/HuangM14,DBLP:conf/cav/CermakLMM14,DBLP:conf/aaai/CermakLM15,DBLP:conf/atva/GutierrezNPW18,DBLP:conf/atal/KurpiewskiPJK21,DBLP:journals/sttt/LomuscioQR17}.  In parallel, temporal and strategic logics were extended to quantitative settings: finite‐trace \LTL\(_F\) with averaging semantics \cite{almagor2016formally}, threshold‐based \CTL\(_F\) \cite{DBLP:journals/fss/MaLLHH24}, continuous‐truth \ATLF \cite{DBLP:conf/prima/FerrandoLMM24}, payoff‐oriented Strategy Logic (\SLF) \cite{bouyer2023reasoning}, as well as timed extensions \cite{BLMO07,HP06}, multi‐valued semantics \cite{JamrogaKKP20,BFM23}, weighted and discounted objectives \cite{DBLP:conf/atal/AminofMMR16,BullingG22,LMO06,Ves15,AlmagorBK14,henzinger2005,MittelmannMP23}, probabilistic frameworks \cite{BJMM24,CL07a,AKMMR19,CAFL09}, and resource‐constrained strategies \cite{NALR18,DBLP:conf/prima/Catta24}. Separately, the paradigm of natural strategies bounded the complexity of the strategy through simple representations based on rules, to improve interpretability without losing the determinability \cite{JMM19a}. Foundational resource logics have also tracked consumable budgets in plays \cite{10.1007/978-3-642-04893-7_1,DBLP:journals/logcom/AlechinaLNR11,DBLP:conf/atal/AlechinaLNRM15}, but none of these approaches is suitable for a more human-like representation of strategies to catch fuzzy achievements of goals using non-refillable resource budgets. 

\paragraph{Outline.}  
Section 2 defines the background notions for the reader. Section 3 provides the syntax and semantics of \HumanATLF. Section 4 presents the model‐checking algorithms analysis. Section 5 describes the implementation. Section 6 reports the evaluation with a drone‐rescue case study. Section 7 concludes the paper. All proofs are provided in the Technical Appendix of the supplementary material.

\section{\uppercase{preliminaries}}
\label{sec:preliminaries}
We begin by introducing resource-bounded fuzzy  Concurrent Game Structures (\rfCGS), fuzzy operators, and the foundation for interpreting \HumanATLF\ formulas and for defining its model checking problem.

\subsection{Resource-Bounded Fuzzy Concurrent Game Structures}

A Concurrent Game Structure (\CGS) \cite{alur2002alternating} is a formal model that effectively represents systems where multiple agents interact with each other and their environment. It is a graph where nodes denote system states, and edges show possible transitions between states. These transitions are non‐deterministic, meaning that agents can choose from multiple possible actions in a given state and each chosen action leads to a new state. The overall outcome of the game emerges from the collective agents choices. Building on this foundation, our field often addresses the inadequacy of the \CGS in verifying quantitative goals by employing a Weighted Concurrent Game Structure (\wCGS) \cite{bouyer2023reasoning, 10.5555/3535850.3535859}. In a \wCGS, every state is further characterized by fuzzy truth values (weights) assigned to propositions, quantifying the value of each property within the game. It is worth to note that, we extend the \wCGS with a resource function $res$ and a $consume$ function to handle action costs. We refer to the resulting model as a resource-bounded fuzzy Concurrent Game Structure (\rfCGS). 

\begin{definition}
	A \rfCGS is a tuple $G=(\Ag, \APf, \{\Act_a\}_{a \in \Ag}, \setpos, \pos_\init, \val, \allowbreak \availability,\transCGS, res, consume)$ where:
	\begin{itemize}
        \item $\Ag$ is a not empty finite set of agents.
        \item $\APf$ is a not empty finite set of atomic propositions (atoms).
		\item For every $a \in \Ag$, $Act_a$ is a not empty finite set of {\em actions}. Let $Act = \bigcup_{a \in \Ag} Act_a$ be the set of all actions, and $ACT = \prod_{a \in \Ag} Act_a$ the set of all joint actions.
		\item $\setpos$ is a finite set of states.
		\item $\pos_\init\in\setpos$ is an initial state.
		\item $\val:\setpos\times\APf\to [\lowb,1]$ is a weight function.
		\item $\availability:\Ag \times \setpos \to 2^\Act$ is an availability function that defines a non-empty set of actions available to agents at each state.
		\item $\transCGS$ is a transition function which assigns the outcome state $\pos' = \transCGS(\pos,\profile{\act})$ to each state $\pos$ and tuple of actions $\profile{\act} \in \prod_{\ag \in \Ag}\availability(\ag,\pos)$ that can be executed by the agents in $\pos$.
        \item $res: \Ag \rightarrow \mathbb{N}$ is the resource function.
        \item $consume:\Ag \times \Act \rightarrow \mathbb{N}$ is the consume function.
	\end{itemize}
\end{definition}
Obviously, $res$ defines the maximum total action cost that agent $a$ can incur during execution. With $consume(a, \alpha)$ we specify the cost for agent $a$ to perform action $\alpha$.
Given a joint action $\profile{\act}$ and a coalition of agents $A$, we use $\profile{\act}_A$ to denote the projection of $\profile{\act}$ onto the actions of the agents in $A$, and $\profile{\act}_{\Ag \setminus A}$ to denote the projection of $\profile{\act}$ onto the actions of the agents outside $A$.

\subsection{Fuzzy Operators}
We characterize the Boolean operators $\land$, $\lor$ and $\neg$ with the quantitative counterparts functions $min\{x,y\}$, $max\{x,y\}$, and $1 - x$. Specifically, $\varphi_1 \lor \varphi_2$ yields the maximum of the values of $\varphi_1$ and $\varphi_2$, $\varphi_1 \land \varphi_2$ yields their minimum, and $\neg \varphi$ equals 1 minus the value of $\varphi$; consequently, the implication $\varphi_1 \rightarrow \varphi_2$ is defined as the maximum of $\varphi_2$ and $(1 - \varphi_1)$. These three truth functions correspond to the original fuzzy logic semantics formulated by Joseph Goguen \cite{goguen1969logic}.

\subsection{Natural Strategies}
We now review the definition of natural strategies and their complexities \cite{JMM19a}, introducing memoryless and recall-based strategies.

\paragraph{Memoryless Strategies.}
We introduce the notion of a \emph{natural memoryless strategy} ($nr$-strategy) $s_a$ for an agent $a$, defined as a rule-based, condition-action representation. Formally, a natural strategy is given by an ordered list of guarded actions, i.e., pairs $(\phi_i,\alpha_i)$ such that:  
\begin{enumerate}
    \item $\phi_i \in \Bool(\APf)$ is a propositional condition on states of the \rfCGS,
    \item $\alpha_i \in d_a(q)$ for every state $q \in \setpos$ where $q \models \phi_i$.
\end{enumerate}
We require that the last pair is of the form $(\top,\alpha)$, ensuring a default action is always available. The collection of all natural memoryless strategies for agent $a$ is denoted by $\Sigma^{nr}_a$. We denote by $length(s_a)$ the number of guarded actions in $s_a$, by $cond_i(s_a)$ and $act_i(s_a)$ the $i$th condition and action respectively, and by $match(q,s_a)$ the smallest index $i \le length(s_a)$ for which $q \models cond_i(s_a)$ and $act_i(s_a) \in d_a(st)$. Moreover, we define $dom(\varphi) = \{ p \in \APf \mid p \in \varphi\}$ and $dom(s_a) = \bigcup_{i=1}^{length(s_a)} dom(cond_i(s_a))$. A \emph{collective natural strategy} for a group of agents $A = \{a_1,\dots,a_{|A|}\}$ is the tuple $s_A = (s_{a_1},\dots,s_{a_{|A|}})$, with the set of such strategies denoted by $\Sigma^{nr}_A$. The outcome function $out(st,s_A)$ returns all paths that can occur when the agents in $A$ execute strategy $s_A$ starting from state $st$. Formally, for $st \in \setpos$:
\begingroup
\footnotesize
\begin{align*}
    & \hspace{-0.3cm} out(st,s_A) = \{\pi \in \Pi \mid \pi[0] = st \land \forall_{i\ge 0}\, \exists_{\alpha_1,\dots,\alpha_{|\Agt|}} : \\
    & \hspace{-0.3cm} \quad (a \in A \Rightarrow \alpha_a = act_{match(\pi[i],s_a)}(s_a)) \land \\
    & \hspace{-0.3cm} \quad (a \notin A \Rightarrow \alpha_a \in d_a(\pi[i])) \land \\
    & \hspace{-0.3cm} \quad (\pi[i+1] = t(\pi[i], \alpha_1,\dots, \alpha_{|\Agt|})) \}.
\end{align*}
\endgroup
Note that $out(st,s_A)$ encompasses \emph{all} paths consistent with $s_A$, without imposing any assumptions on the strategies or behavior of the opponents. 

\paragraph{Recall-based Strategies.}
Memoryless strategies are often insufficient when decisions depend on the history of the game. Agents with memory can base their choices on past states, which can be represented using automaton states \cite{vester2013alternating}. However, we propose a more intuitive approach: using regular expressions over propositional formulas. So, let $Reg(L)$ be the set of regular expressions over the language $L$ with standard operations concatenation $\cdot$, nondeterministic choice $\bigcup$, and Kleene star $\ast$. A natural strategy with recall ($nR$‐strategy) $s_{a}$ for agent $a$ is a sequence of pairs from $Reg(\Bool(\APf)) \times \Act$, namely a pair $(r, a)$ with $r$ being a regular expression over $\Bool(\APf)$, and $a$ an action available in $\last(h)$ for all histories $h$ consistent with $r$. Formally, given a $r$ and the language $L(r)$ on words generated by $r$, an $h=q_0 \dots q_n$ is consistent with $r$ iff $\exists \: d \in L(r)$ such that $|h| = |d|$ and $\forall_{0 \leq j \leq n} \: h[j] \models d[j]$. For an individual agent $a$, the set of all such strategies is denoted $\Sigma^{nR}_a$. To decide which rule applies to a given $h$ the matching function $match(h[0,j],s_{a})$ is the smallest $n\le length(s_a)$ such that $\forall_{0 \leq m \leq j} h[j] \models cond_n(s_a)[m]$ and $act_n(s_a) \in d_a(\pi[m])$. The function $out(q, s_a)$ continues to denote the set of all paths starting from a state $q$ that are consistent with the collective strategy $s_a$. 

\paragraph{Natural Strategy Complexity}
The \textit{complexity} $compl(s_{a})$ of a natural strategy is determined by the size of its representation. In the case of $nr$-strategies, complexity is measured by the total number of symbols in the Boolean conditions, while for $nR$-strategies, it is quantified by the total size of all the regular expressions used. A collective strategy ($s_{\coalition}$) is equal to the sum of its individual strategies complexities ($s_{a}$), formally $compl_{\Sigma(s_{A})}=\sum_{i=1,...,n}compl_{\Sigma}(s_{a_{i}})$. Hereafter in this paper, we will refer to each strategy type using the abbreviations $nr$ and $nR$. By adopting natural strategies, we establish \HumanATLF\ combining them with a quantitative approach.
\paragraph{Model Path.}
In a \rfCGS $G$, a path $\pi$ represents an infinite sequence of states. The set of paths over $St$ is denoted by $St^\omega$. For a joint natural strategy $\sigma_\Gamma$, consisting of one strategy for each agent in coalition $\Gamma$, a path $\pi$ is $\sigma_\Gamma$-compatible if, for every $j \geq 1$, $\pi_{j+1} = \text{out}(\pi_j, c)$ for some joint action $c$ such that for every $i \in \Gamma$, $c_i = \sigma_i(\pi_{\leq j})$, and for every $i \in \Gamma$, $c_i \in d(i, \pi_j)$. The set of all $\sigma_\Gamma$-compatible paths from $s$ is denoted by $\text{out}(s,\sigma_\Gamma)$.

\section{\uppercase{\HumanATLF}}
\label{sec:human}
In this section, we provide a more precise formalization of human-like strategies, that takes into account a cost over the actions and then give the formal syntax and semantics of \HumanATLF and illustrate its use on the concrete example drawn from the aforementioned drone scenario.

\paragraph{Human Resource-Bounded Strategies.}
We extend the definition of natural strategies by incorporating resource constraints on the feasibility of actions. As stated above, a natural strategy for an agent $a$ is a sequence of guarded actions $(\varphi, \alpha)$. We define cost over action integrating it as $(\varphi, \alpha_\tau)$, where $\varphi$ is a Boolean formula over atomic propositions, defining the guard under which the rule applies; $\alpha \in d(a, s)$ is an action available to agent $a$ in any state $st$ such that $\CGS, st \models \varphi$; $\tau \in \mathbb{N}$ is the fixed resource cost for executing action $\alpha$. The association of costs and actions are defined via the resource consumption function $consume : \Ag \times \Act \rightarrow \mathbb{N}$, where $consume(a, \alpha)$ specifies the cost for agent $a$ to perform action $\alpha$. Each rule in a natural strategy must satisfy $\tau = consume(a, \alpha)$, ensuring that costs are consistently assigned based solely on the agent and the action, independent of the system state. Each agent $a \in \Ag$ is initially endowed with a finite quantity of resources, given by the function: $res : \Ag \rightarrow \mathbb{N}$, which defines the maximum total cost that agent $a$ can incur during actions. In particular, for each rule $(\varphi,\alpha_\tau)$ in the strategy of agent $a$, it must hold that the collection of total action cost $b \leq res(a)$. Before the presentation of \HumanATLF, we introduce some notation that will be used throughout the paper. We denote the length of a tuple $v$ as $|v|$, its $j$-th element as $v_j$, and its last element $v_{|v|}$ as $last(v)$. 
For $j \leq |v|$, let $v_{\geq j}$ be the suffix $v_{j},\ldots, v_{|v|}$ of $v$ starting from $v_j$ and $v_{\leq j}$ the prefix $v_{1},\ldots, v_{j}$ of $v$.

\paragraph{\HumanATLF Syntax.}
The grammar of \HumanATLF\ is given by the following definition:
\begin{definition}
\label{def:NatATLf}
Formulas $\varphi$ in \HumanATLF are defined as follows:

\begin{center}
    \scriptsize
    $\varphi ::= p \mid f[\varphi, \ldots, \varphi] \mid \coop{\coalition}^{\scriptscriptstyle\leq k}_{\scriptscriptstyle\mkern2mu\leq b} \X \varphi \mid \coop{\coalition}^{\scriptscriptstyle\leq k}_{\scriptscriptstyle\mkern2mu\leq b} (\varphi \until \varphi) \mid \coop{\Gamma}^{\scriptscriptstyle\leq k}_{\scriptscriptstyle\mkern2mu\leq b} (\varphi_1 \release \varphi_2)$
\end{center}

where $p\in\APf$, $\coalition\in2^\Ag$, $k \in \mathbb{N}$ is the complexity bound for the strategies of agents in $\coalition$, $b \in \mathbb{N}$ is the resource bound over actions, and $f\in \Func$, where $\Func \subseteq \{ f: [0,1]^m \rightarrow [0,1] \ | \ m \in \mathbb{N} \}$ represents the set of computable functions. Temporal operators \textbf{G} and \textbf{F} are classically derived.
\end{definition}

\paragraph{\HumanATLF Semantics.}
Formulas of \HumanATLF\ are evaluated over a weighted \rfCGS $G=(S, \dots,\val,\dots)$ with atomic propositions in $[0,1]$. Let $\pi=(st_{1},st_{2},\dots)$ be a path. The satisfaction degree $\semG{\varphi}{\pi}$ is defined by:

{\scriptsize
\begin{itemize}[leftmargin=*, labelsep=0.3em, itemsep=0pt, parsep=0pt, topsep=0pt]
  \item $\semG{p}{\pi} \;=\; \val(st_{1},p)$
  \item $\semG{f[\varphi_{1},\dots,\varphi_{m}]}{\pi} 
        = f\bigl(\semG{\varphi_{1}}{\pi},\dots,\semG{\varphi_{m}}{\pi}\bigr)$
  \item $\displaystyle
        \semG{\coop{\Gamma}^{\le k}_{\le b}\X\psi}{\pi}
        =\max_{\sigma\in\Sigma^{k,b}_{\Gamma}}(
         \min_{\pi'\in\out(st_{1},\sigma)}(
         \semG{\psi}{\pi'_{\ge2}}))$
  \item $\displaystyle
        \semG{\coop{\Gamma}^{\le k}_{\le b}\G\psi}{\pi}
        =\max_{\sigma\in\Sigma^{k,b}_{\Gamma}}(
         \min_{\pi'\in\out(st_{1},\sigma)}(
         \min_{j\ge1}(\,\semG{\psi}{\pi'_{\ge j}})))$
  \item $\displaystyle
        \semG{\coop{\Gamma}^{\le k}_{\le b}(\psi_{1}\U\psi_{2})}{\pi}
        =\\ = \max_{\sigma\in\Sigma^{k,b}_{\Gamma}}(
         \min_{\pi'\in\out(st_{1},\sigma)}(
         \max_{j\ge1}(\,\min\!\bigl(\semG{\psi_{2}}{\pi'_{\ge j}},\min_{i<j}(\semG{\psi_{1}}{\pi'_{\ge i}}\bigr)$
  \item $\displaystyle
        \semG{\coop{\Gamma}^{\le k}_{\le b}(\psi_{1}\R\psi_{2})}{\pi}
        =\\ = \max_{\sigma\in\Sigma^{k,b}_{\Gamma}}(
         \min_{\pi'\in\out(st_{1},\sigma)}(
         \min_{j\ge1}(\,\max\!\bigl(\semG{\psi_{2}}{\pi'_{\ge j}},\max_{i<j}(\semG{\psi_{1}}{\pi'_{\ge i}}\bigr)$
\end{itemize}
}  

$\Sigma^{k,b}_{\Gamma}$ is the set of all $\Gamma$–natural strategies with complexity at most $k$ and total actions cost at most $b$, and $\out(st_{1},\sigma)$ denotes the outcome paths from $st_{1}$ under strategy $\sigma$.

\begin{figure}[h]
    \centering
    \includegraphics[width=180pt]{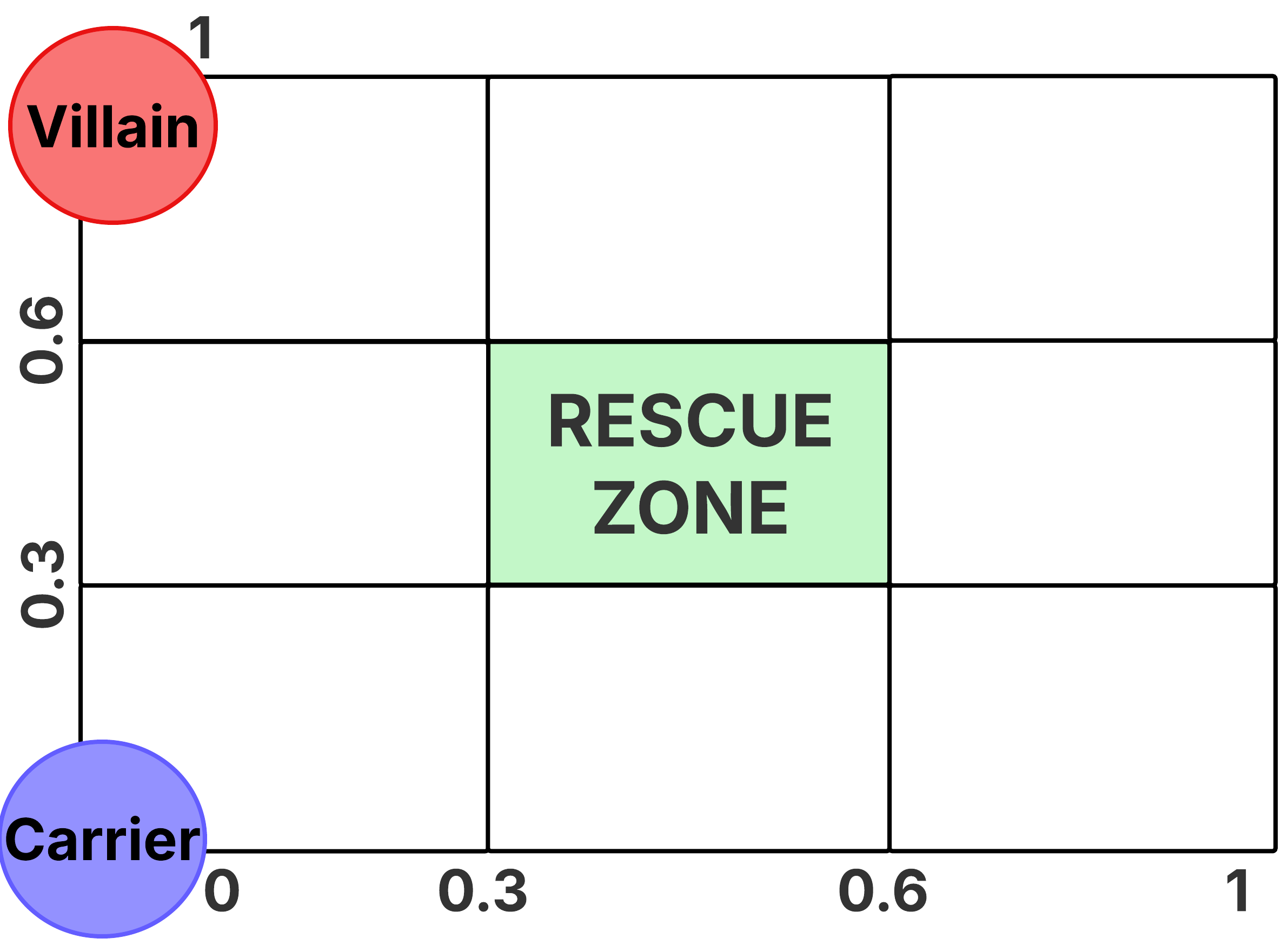}
    \caption{\footnotesize Drone Battle scenario: the carrier and villain occupy grid positions and evaluate \texttt{dist} and \texttt{safe} on each state.}
    \label{fig:table}
\end{figure}

\begin{example}
\label{ex:drone_coalition}
Consider the coalition \(\coop{\mathit{carrier},\mathit{drone}}\) operating on the model from Figure \ref{fig:table} at the carrier position pawn, with strategy complexity bound \(k = 2\), the total energy budget \(b = 5\), and the atomic propositions
\[
  p \;=\;\mathit{safe}, 
  \quad
  q \;=\; (\mathit{dist}\le0.5),
\]
where \(p\) holds in the green “Rescue Zone” and \(q\) flags any state whose carrier–villain distance drops below \(0.5\). We have the \(\HumanATLF\) formula
\[
  \varphi \;=\;
  \coop{\mathit{carrier},\mathit{drone}}^{\le2}_{\le5}(\,\neg q\;\until\;p\,),
\]
meaning the coalition has a natural strategy of at most two guard‐action rules and spends no more than five energy units in total, so that the carrier reaches the Rescue Zone while the villain never comes within \(0.5\).

A suitable joint strategy of complexity 2 is:

\[
\small
\begin{aligned}
\textbf{Rule 1:}\quad 
&\textbf{if }(x\le0.6 \wedge y\ge0.3)\textbf{ then}\\
&\quad(\mathit{carrier}:\,\mathit{move\_right},\;\mathit{drone}:\,\mathit{move\_right}),\\[4pt]
\textbf{Rule 2:}\quad
&\textbf{otherwise if }(x>0.6)\textbf{ then}\\
&\quad(\mathit{carrier}:\,\mathit{ascend},\;\mathit{drone}:\,\mathit{ascend}).
\end{aligned}
\]

Here each agent’s action costs \(1\) energy unit per step, so a joint step costs \(2\).  Starting from \((0,0)\), the coalition (i) executes Rule 1 twice, moving both agents to column \(x=0.6\) (cost \(2\times2=4\)), then (ii) applies Rule 2 once to ascend into the Rescue Zone \([0.3,0.6]\times[0.3,0.6]\) (cost 2), for a total cost of \(6\).  To respect the budget \(b=5\), one can instead perform a single application of Rule 1 (cost 2) followed by a single application of Rule 2 (cost 2), then a final move by the carrier alone (cost 1) while the drone hovers (still described by Rule 2) bringing the total to \(2+2+1=5\).  Throughout, \(\mathit{dist}>0.5\) is maintained, so \(\varphi\) is satisfied.
\end{example}

\section{\uppercase{Model Checking}}
\label{sec:modelchecking}

In this section, we study algorithms and the complexity of the model-checking problem for \HumanATLF with both \textit{nr}-strategies and \textit{nR}-strategies, i.e., \HumanATLrF and \HumanATLRF. We analyze both constant and variable bounds k on the size of natural strategies. 

\subsection{\HumanATLrF Model Checking}
In this subsection, we establish that the model checking problem for \HumanATLrF remains computationally efficient when the complexity bound $k$ and the resource bound $b$ are fixed. The model checking algorithm runs in linear time with respect to the size of the model, and the overall complexity remains polynomial in the size of the model and the length of the formula. The formal result follows.

\begin{theorem}\label{the1}
     The model checking problem for \HumanATLrF with fixed complexity bound $k$ and fixed resource bound $b$ is in $P$ with respect to the size of the model and the length of the formula.
\end{theorem}

\begin{proof}
Let $\varphi = \coop{\coalition}^{\leq k}_{\leq b} \psi$ be a \HumanATLrF formula, where $\psi$ is built using Boolean connectives, atomic propositions, and the weight function $\val$. We assume that any collective strategy $\sigma_{\coalition}$ available to coalition $\coalition$ is bounded by $k$ and incurs action cost at most $b$. Since natural strategies are built from guarded actions whose guards are Boolean formulas over atomic propositions with constant weights provided by the function $\val$, the number of possible collective strategies is at most $O((|\Prop| + |\Bool|)^{k^2} \cdot |\Act|^k)$, where the constants from $\val$ are absorbed in the bound. For each such strategy $\sigma_{\coalition}$, we perform a pruning step on the input \wCGS. That is, we remove every transition that does not agree with the moves prescribed by $\sigma_{\coalition}$. This pruning takes $O(|t|)$ time, where $t$ is the transition relation of the \wCGS. We then add a resource feasibility filter: for each strategy, we check whether its cumulative cost remains within the resource bound $b$ along all possible outcome paths. This check can be done in linear time, and in the worst case, its cost is proportional to the size of the strategy itself. Only strategies that pass this feasibility check are retained. After pruning and filtering, the resulting structure can be viewed as a weighted Kripke structure. The model checking task now reduces to verifying a quantitative \CTLF formula. In \CTLF, the universal path quantifier is written as $\forall \psi$, where its semantics is defined as the minimum of the truth values of $\psi$ over all paths from a state $s$: $\llbracket \forall \psi \rrbracket(s) = \min_{\pi \in \Pi(s)} \{ \llbracket \psi \rrbracket(\pi) \}$. Model checking for \CTLF is known to be linear in the size of the structure and the formula \cite{DBLP:journals/fss/MaLLHH24}. Thus, the total complexity is $O\big((|\Prop| + |\Bool|)^{k^2} \cdot |\Act|^k \cdot |t| \cdot |\psi|\big)$, which is polynomial in the size of the model and the formula, assuming $k$ and $b$ are fixed. Finally, since a formula may contain multiple strategic operators, we adopt a bottom-up evaluation strategy: we recursively resolve the innermost strategic operators, update both the formula and the structure accordingly, and repeat until all operators are evaluated.
\end{proof}

We provide below an algorithm that summarizes the verification procedure described in the proof of Theorem~\ref{the1}.
\begin{algorithm}[ht]
\captionsetup{font=scriptsize}
\caption{HumanATL[F] Memoryless Model Checking}
\label{alg:NatATLrF_Handler}
\scriptsize
\begin{algorithmic}[1]
    \Procedure{HumanATLFMemorylessModelCheck}{model, $\coop{A}^{\leq k}_{\leq b}\psi$}
        \ForAll{ $\sigma$ in GenerateStrategyCandidates(model, $A$, $k$, $b$)}
            \State $M_{\sigma} \gets \text{PruneModel}(model, \sigma)$
            \If{CTL[F]ModelChecking($M_{\sigma}$, $\psi$)}
                \State \Return TRUE
            \EndIf
        \EndFor
        \State \Return FALSE
    \EndProcedure
\end{algorithmic}
\end{algorithm}

We now examine the complexity of \HumanATLrF when the bounds in the strategic modalities are treated as variables. We begin by establishing NP-completeness for formulas that consist of a single strategic operator followed by a simple temporal subformula. Then, we modify the argument to demonstrate that model checking for the entire \HumanATLrF is $\Delta^P_2$-complete.

\begin{proposition}\label{the2}
 The model  check for $1\HumanATLrF$ (i.e., with a single strategic operator) is in NP with respect to the size of the model, the length of the formula, the variable value of the bound $k$, and the variable value of the bound $b$.
\end{proposition}

\begin{proof}
Consider a formula of the form $\varphi = \coop{A}^{\leq k}_{\leq b}\gamma$, where $A \subseteq \Agt$ and $\gamma$ is a temporal formula with a single temporal operator. Instead of exhaustively enumerating all appropriate strategies and checking each one, we non-deterministically guess a collective strategy $s_A$ and proceed with an argument analogous to that in the proof of Theorem~\ref{the1}. Since the size of $s_A$ is polynomial in the model size, the algorithm remains in NP.
\end{proof}

An algorithm that represents the proof is described in Algorithm \ref{alg:1ResNatATLrF_NP}. We emphasize that this result is stated in terms of the value of $k$ and $b$, or equivalently, the size of its unary encoding. This perspective will be maintained throughout the paper, and the analysis of model-checking complexity with respect to the binary representation of $k$ and $b$ is left for future research.

\begin{algorithm}[H]
\captionsetup{font=scriptsize}
\caption{$1\HumanATLrF$ Model Checking}
\label{alg:1ResNatATLrF_NP}
\scriptsize
\begin{algorithmic}[1]
    \Procedure{1HumanATLFModelChecking}{$M,\,\coop{A}^{\leq k}_{\leq b} \gamma$}
        \Guess{$s_A$ in \text{GenerateStrategyCandidates{($M, A, k, b$)}}}
            \State $M_{\sigma} \gets \text{PruneModel}(model, \sigma)$
            \If{CTL[F]ModelChecking($M_{\sigma}$, $\psi$)}
                \State \Return TRUE
            \EndIf
        \EndGuess
        \State \Return FALSE
    \EndProcedure
\end{algorithmic}
\end{algorithm}

\begin{theorem}\label{Theorem2}
    $1\HumanATLrF$ is NP-complete with respect to the size of the model, the length of the formula, and the variable value of k.
\end{theorem}

\begin{proof}
The lower bound follows from~\cite{JMM19a}, since $1\NatATLr$ is an NP-hard problem. The lower bound is not affected by resources and action costs, as these are requirements that restrict the strategy space but do not alter the hardness. 
The upper bound follows by Proposition~\ref{the2}.
\end{proof}

Given the above result, we can present the general case.

\begin{theorem}\label{Theorem3}
Model Checking \HumanATLrF is $\Delta^P_2$-complete with respect to the size of the
 model, the length of the formula, the maximal bound $k$, and maximal cost $b$.
\end{theorem}

\begin{proof}
We establish the upper bound by structural induction on the number of nested strategic operators. For the innermost formula of the form $\langle\langle A \rangle\rangle^{\leq k}_{\leq b} \psi$, we non-deterministically guess a natural strategy $s_A$ of bounded complexity $k$ and total cost at most $b$. This guess can be verified in polynomial time via pruning and feasibility filtering as shown in Proposition~1. To handle nested modalities, we evaluate subformulas recursively: each strategic operator introduces a new non-deterministic guess of a strategy, followed by a coNP check of universal path satisfaction, and a polynomial-time fixpoint evaluation for temporal constructs. Consequently, the model-checking problem is solvable in deterministic polynomial time with access to an NP oracle, placing it in $\Delta^P_2$. For the lower bound, since $\HumanATLrF$ subsumes $\NatATLr$~\cite{JMM19a}, the result carries over directly.
\end{proof}

\subsection{\HumanATLRF Model Checking}
We now examine the \HumanATLRF model-checking problem. Bounded tree unfoldings represent outcome sets of recall strategies.

\begin{proposition}\label{propositionR}
Let \(M\) be a \rfCGS with a finite state set \(\mathrm{St}_M\), and let 
$s_A = (s_{a_1}, \dots, s_{a_n}) \in \Sigma^{nR}_A$ be a natural resource-bounded strategy with recall for coalition \(A\) with complexity $k = \mathrm{compl}(s_A)$. Suppose that every action has an associated cost, and each agent starts with a fixed, non-refillable amount of resource $res(a) \in \mathbb{N}$. Assume that goal is a temporal formula of the form \(\varphi\,\mathcal{U}\,\psi\). Then, to decide whether \(s_A\) enforces the objective from a state \(st \in \mathrm{St}_M\), it is sufficient to consider the tree unfolding of the executions prescribed by \(s_A\) up to a depth $L = |\mathrm{St}_M| \cdot 2^{2k^2} \cdot \prod_{a \in A} (r_a + 1)$.

\end{proposition}

\begin{proof}
In the automata-based approach, each regular-expression condition in the nR-strategy $s_A$ is first converted into a deterministic finite automaton (DFA). A regular expression of length at most $k$ can, in the worst-case, be converted into a DFA with at most $2^{2k}$ states. This bound is obtained by considering that the nondeterministic automaton derived from the regular expression may have $O(k)$ states, and applying the powerset construction yields a DFA with at most $2^{O(k)}$ states-in our estimate we use the bound $2^{2k}$. Since the strategy $s_A$ comprises a number of such conditions whose total complexity is bounded by $k$, the combined state information obtained by considering all these DFAs is given by the product of the number of states of each DFA. In the worst-case, if there are roughly $k$ such conditions, the total number of distinct combinations is bounded by 
$\left(2^{2k}\right)^k = 2^{2k^2}$. A configuration in the execution tree is given by a pair consisting of the current state of $M$ and the current states of the DFAs corresponding to the regex conditions in $s_A$. Since the state space \(\mathrm{St}_M\) is finite, the total number of distinct configurations is at most $|\mathrm{St}_M| \cdot 2^{2k^2}$. To account for the additional constraints imposed by resource-bounded actions, we must also track the current resource availability of each agent in the coalition. Since resources are non-refillable and bounded by the initial values \(r_a\), each agent \(a \in A\) can have \(r_a + 1\) distinct remaining resource levels (ranging from 0 to \(r_a\)). The total number of resource configurations across all agents is therefore bounded by the product \(\prod_{a \in A} (r_a + 1)\). Accordingly, the total number of distinct configurations is at most
\[
|\mathrm{St}_M| \cdot 2^{2k^2} \cdot \prod_{a \in A} (r_a + 1).
\]

Hence, any path in the tree unfolding longer than $L = |\mathrm{St}_M| \cdot 2^{2k^2} \prod_{a \in A} (r_a + 1)$ must repeat a configuration, forming a cycle. As the objective \(\varphi\,\mathcal{U}\,\psi\) is a reachability property (requiring that \(\psi\) eventually holds while \(\varphi\) holds until then), once a cycle is detected, further exploration does not yield additional information regarding the satisfaction of the objective. Therefore, to decide whether \(s_A\) enforces \(\varphi\,\mathcal{U}\,\psi\) from \(q\), it suffices to explore the unfolding up to depth \(L\).
\end{proof}

Based on the above proposition, we can now present the decision procedure as a concrete algorithm.  Algorithm~\ref{alg:BoundedUnfolding} implements the bounded‐depth unfolding up to \(L = |\mathrm{St}_M|\cdot2^{2k^2} \prod_{a \in A} (r_a + 1)\) and checks the \(\varphi\,\mathcal{U}\,\psi\) objective under \(s_A\).

\begin{algorithm}[ht]
\captionsetup{font=scriptsize}
\caption{Bounded‐Depth Unfolding for \(\varphi\,\mathcal{U}\,\psi\) under \(s_A\) with resources}
\label{alg:BoundedUnfolding}
\scriptsize
\begin{algorithmic}[1]
  \Procedure{CheckObjectiveUnderStrategy}{$M, s_A, q, \varphi, \psi, \{r_a\}_{a\in A}$}
    \State // Build DFAs for each regex condition in $s_A$
    \ForAll{regex $r$ in $s_A$}
      \State $D_r \gets \textsc{RegexToDFA}(r)$
    \EndFor
    \State // Compute the combined depth bound including resources
    \State $L \gets |\,\mathrm{St}_M| \times 2^{2k^2} \times \prod_{a\in A}(r_a + 1)$
    \State // Explore the outcome tree up to depth $L$
    \State Initialize queue $Q$; enqueue $(s=q,\; \{q_r = D_r.\mathit{init}\}_{r},\; d=0)$
    \While{$Q$ not empty}
      \State $(s,\,(q_r)_r, d) \gets Q.\text{dequeue}()$
      \If{$d > L$} \State \textbf{continue} \EndIf
      \If{$\neg\varphi(s)$} \State \Return \textbf{FALSE} \EndIf
      \If{$\psi(s)$} \State \textbf{continue} \EndIf
      \ForAll{joint action $m$ prescribed by $s_A$ at $(s,(q_r)_r)$}
        \State $s' \gets \textsc{NextState}(M, s, m)$
        \ForAll{regex $r$}
          \State $q'_r \gets D_r.\delta(q_r, \textsc{Observe}(s,m))$
        \EndFor
        \State Enqueue $(s',\,(q'_r)_r,\; d+1)$ into $Q$
      \EndFor
    \EndWhile
    \State \Return \textbf{TRUE}
  \EndProcedure
\end{algorithmic}
\end{algorithm}

We now establish that model checking for \HumanATLRF can be performed using only polynomial space.

\begin{theorem}
Model Checking of \HumanATLRF is in PSPACE with respect to the size of the model, the formula length, the maximal complexity bound $k$, and the maximal cost bound $b$.
\end{theorem}

\begin{proof}
Let $\varphi = \langle\!\langle A \rangle\!\rangle_{\leq k}^{\leq b} \psi$ be a $\HumanATLRF$ formula over a \wCGS $M$, where $\psi$ is a temporal formula interpreted under fuzzy semantics. The goal is to decide whether there exists a collective natural recall strategy $s_A \in \Sigma^{nR}_A$ of complexity at most $k$ and total cost within $b$, such that all paths in $\mathrm{out}(s_A)$ satisfy $\psi$.By Proposition 2, it suffices to explore the execution tree up to depth $L = |\mathrm{St}_M| \cdot 2^{2k^2} \cdot \prod_{a \in A}(r(a)+1)$, where $res(a)$ is the initial resource of agent $a$. Each configuration in this unfolding encodes a state of $M$, the current states of the DFAs compiled from the strategy's regular expressions, and the residual resources of each agent.This exploration is conducted via depth-first traversal, which requires only polynomial space: each DFA state is representable in $\mathcal{O}(k^2)$, each resource level in $\mathcal{O}(\log b)$, and the overall number of agents and automata is fixed by the input. Strategy candidates can be guessed nondeterministically, and feasibility under cost bounds can be verified incrementally along the path without materializing the full tree. The fuzzy satisfaction of $\psi$ is evaluated compositionally along each branch, and the unfolding-based model-checking procedure follows the approach described in \ref{alg:BoundedUnfolding}. Since each recursive subformula is evaluated over bounded paths using local information only, and nesting is resolved bottom-up, the entire verification proceeds in polynomial space. Thus, the procedure is a nondeterministic polynomial-space algorithm. Since $\textsc{NPSpace} = \textsc{PSpace}$, we conclude that model checking for $\HumanATLRF$ is in $\textsc{PSpace}$.
\end{proof}

\section{Implementation}
\label{sec:Implementation}

\begin{figure}[h]
    \centering
    \includegraphics[width=0.99\linewidth]{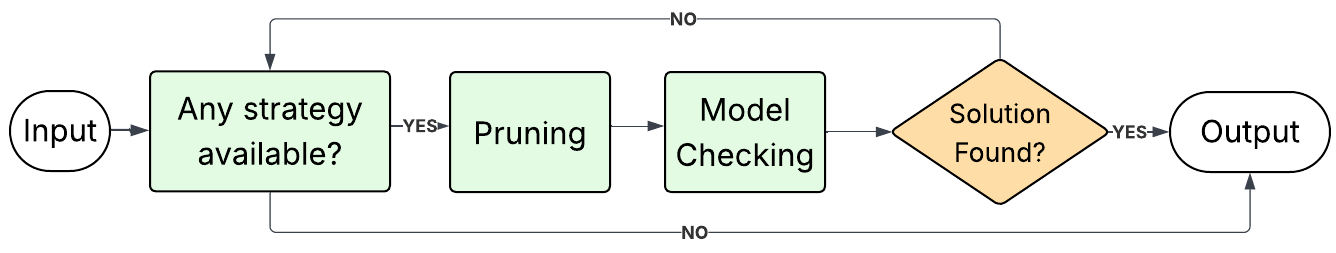}
    \caption{\HumanATLF Model Checking process}
    \label{fig:flowchartNatATLF}
\end{figure}

In this section, we present the algorithms that implement the proposed \HumanATLF logic. The implementation is designed to be consistent with the theoretical foundations established in the previous sections. Figure~\ref{fig:flowchartNatATLF} illustrates our \HumanATLF model‐checking workflow, implemented in Python and organized into three phases: Strategy Generation, Model Pruning, and Model Checking.
In the Strategy Generation phase, we enumerate each agent’s guarded actions by pairing guards (boolean or regex conditions) with available moves, form complete strategy profiles via Cartesian product, and enforce bounded deviation: for every strategy profile, each alternative action must also respect the specified complexity bound. A guarded action $(\varphi,a)$ is accepted only if $\varphi$ holds in all states where $a$ is enabled and the sum of action costs (via \texttt{get\_action\_cost()}) does not exceed the agent’s resource budget $b$. During Model Pruning, we refine the input model for each valid collective strategy $s_A$ by removing all transitions that do not conform to $s_A$, producing a smaller model tailored for efficient verification.  In model checking for \HumanATLF, we adopt a binary abstraction on top of the underlying fuzzy semantics: although \HumanATLF formulas are interpreted over the continuous range $[0,1]$, we introduce meta-truth values \textit{True} and \textit{False} to indicate whether a formula is considered satisfied in a given state. After computing the fuzzy satisfaction degrees, we treat a formula as \textit{True} exactly in those states whose degree meets or exceeds a designated threshold (in our implementation, effectively those target states where the computed degree equals 1.0\footnote{obviously, if the user needs a narrower truth range it possible to associate more fuzzy-values to the true meta value}), and as \textit{False} otherwise. These meta-values do not replace the fuzzy degrees but serve as a practical decision criterion for pruning and strategy validation.  More specifically, the representation of fuzziness in our framework is realized via an explicit fuzzy Kripke structure (FKS) abstraction. From the original input model, we extract a mapping $R: S \times S \to [0,1]$ by assigning to each joint action label a normalized degree in the interval $(0,1)$ - using a simple Gougen fuzzy‐logic scheme where
\[
  \mu(a) = \frac{\mathrm{index}(a)}{|\mathsf{Labels}|+1}\,,
\]
and propagate these degrees into the transition relation. Atomic propositions are also annotated with fuzzy truth values in $V: S \times $\APf$ \to [0,1]$ according to the input labelling. We then employ classical fuzzy‐CTL fixed-point algorithms (EX, AX, EU, EG, AU, AF, AG) instantiated with Gouguen’s minimum and maximum, as well as standard negation and implication operators, to compute a real‐valued satisfaction map 
\(\,[\,\![\varphi]\!] : S \to [0,1]\)\,.  
The degree of satisfiability for each state is then used both to inform the meta-truth abstraction and to drive the decision whether a strategy yields a winning outcome. This integrated approach ensures that our implementation faithfully realizes the fuzzy semantics of \HumanATLF with resource bound actions while maintaining practical efficiency.  A key aspect of our implementation is that, beyond indicating whether a solution exists for the given (pruned) model, the algorithm also synthesizes the optimal winning strategy leading to that solution. Specifically, if a solution validating the formula is found, the tool returns both the set of states that satisfy it and the minimal‐complexity strategy that guarantees its satisfaction. This ensures that, rather than merely verifying existence, our framework actively constructs and returns an executable strategy for the agents. If the verification does not yield a solution for the pruned model, the algorithm backtracks to the Strategy Generation phase, projects new strategies onto the original model, and re-attempts verification. This iterative cycle continues until a valid strategy is synthesized or no further strategies remain viable.

\section{\uppercase{Experiments}}
\label{sec:Experiments}
We evaluated our \HumanATLF verifier on an HPOmen 15‑ax213ng (Intel i7‑7700HQ 3.8GHz, 16GB RAM), implemented in Python 3.9 (PyCharm) and integrated into the open‑source VITAMIN checker \cite{DBLP:conf/eumas/Ferrando24}. VITAMIN provides a robust baseline, supporting memoryless and bounded‑recall strategies with bounded complexity, against which we compared our module. Unlike MCMAS \cite{DBLP:journals/sttt/LomuscioQR17}, its implementation handles natural strategies. To validate correctness, we ran 100 tests per temporal operator, using identical inputs for both \ATLF and our \HumanATLF formulas, with a complexity bound up to 5. Across nearly 1000 trials, \HumanATLF matched the established \ATLF results, confirming implementation fidelity. For performance, we conducted 1000 additional tests, varying model size (number of states), complexity bound, resource bound, and agent count. Execution time grew with both the number of agents and the complexity bound, since each agent’s strategy set expands with its action space. For bounded‑recall strategies, we fixed the tree depth at 5 to avoid excessive unwind time\footnote{Depths $>$5 caused noticeable delays during tree traversal.}. Recall‑based histories added overhead compared to memoryless runs.

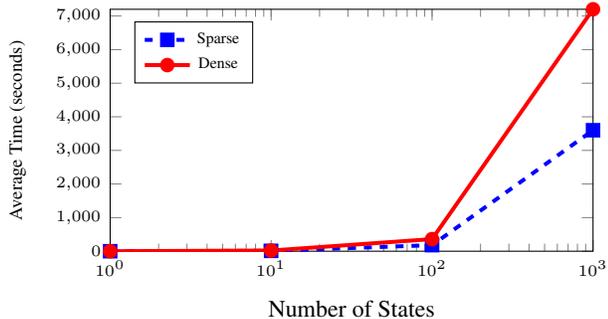
\begin{figure}[ht]
\centering
\begin{minipage}[t]{0.45\textwidth}
\centering
\begin{tikzpicture}
\begin{axis}[
    ylabel style={font=\scriptsize},
    xlabel={Number of States},
    ylabel={Average Time (seconds)},
    xlabel style={font=\small},
    xmode=log,
    log basis x=10,
    xmin=1, xmax=1000,
    ymin=0, ymax=7200,
    ytick={0, 1000, 2000, 3000, 4000, 5000, 6000, 7000}, 
    xtick style={font=\tiny},
    ytick style={font=\tiny},
    tick label style={font=\tiny},
    width=\textwidth, height=0.6\textwidth,
    legend style={font=\tiny, fill=none, at={(0.05,0.95)}, anchor=north west},
    ]
\addplot[blue,dashed,mark=square*,mark options={solid},line width=1.5pt] coordinates {
    (1,0)
    (10,10.58) 
    (100,180) 
    (1000,3600) 
};
\addplot[red,solid,mark=*,line width=1.5pt] coordinates {
    (1,0)
    (10,20.12) 
    (100,360) 
    (1000,7200)
};
\legend{Sparse, Dense}
\end{axis}
\end{tikzpicture}
\caption{Average case: Average time table comparison using dense and sparse transition matrix and $|A|=3$}
\label{table3.1}
\end{minipage}
\hfill
\end{figure}

Figure~\ref{table3.1} illustrates average‑case execution times on sparse vs.\ dense transition matrices with three agents, k=5, and a varying resource bound (as it does not significantly impact performance under our implementation). Dense matrices (higher branching) incur substantially longer runtimes, while sparse matrices execute more quickly. We additionally measured worst‑case behavior on fully nondeterministic models (maximally dense matrices) and average‑case behavior on classical incomplete multi‑agent models, observing that filtering invalid strategies yields more efficient, typical runtimes. 

\subsection{Case Study: Resource‐Bound Drone Battle}

To demonstrate the practical expressiveness of \HumanATLF, we consider a case study grounded in autonomous aerial navigation, a domain where strategy design must inherently balance correctness, readability, and resource feasibility. Our scenario is set in an war-like environment where battery recharging is impossible, emphasizing the critical importance of judicious energy management.  In our example, a carrier drone is tasked with transporting a critical artifact to a designated rescue zone while continuously avoiding proximity to a hostile pursuer (the villain drone). The spatial environment is discretized along both axes into the finite domain $\{0, 0.3, 0.6, 1\}$, effectively modeling a grid-based operational area. Each drone can move in discrete steps, consuming irreplaceable battery power at each action. The environment is formally encoded as a \rfCGS, enriched with action costs and fuzzy valued propositions such as \texttt{dist} (distance between drones) and \texttt{safe} (degrees of positional security), both ranging over $[0,1]$.

Each guarded rule in a strategy not only specifies a boolean or regex condition, but also consumes energy when its action is done. For instance, an “ascend” maneuver might cost 2 battery units, “right” costs 3, and “idle” costs 1. Before pruning, every candidate strategy is checked so that the sum of its per‐use action costs does not exceed the drone’s battery budget (e.g.\ $b=5$). Only those strategies that both satisfy their distance‐based conditions and remain within budget pass to the pruning phase. Strategically, the carrier uses two guarded rules ($k=2$) to avoid the villain and reach safety:
$
  s_{\mathrm{carrier}} \;=\;\{(\neg(\,\mathrm{dist}<0.5\,),\rightarrow),\;(\top,\uparrow)\}
  $.
  
The first rule tells the carrier to move right when the villain is too close, the second to ascend otherwise; the combined energy cost along the path (e.g.\ $3 + 2 = 5$) exactly matches the budget. The villain’s one‐rule strategy ($k=1$) with $b=4$ is specified as
$
  s_{\mathrm{villain}} \;=\;\{(\mathrm{dist}\ge0.5,\downarrow)\},
$
pursuing the carrier whenever distance permits.
\begin{figure}[h]
    \centering
    \includegraphics[width=200pt]{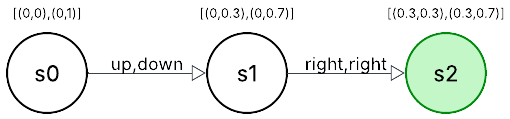}
    \caption{\footnotesize Pruned Drone Battle model: shows only transitions whose joint actions meet both strategic guards and energy budgets.}
    \label{fig:graph}
\end{figure}
After pruning, the resulting game graph (Figure~\ref{fig:graph}) retains only states reachable under the selected strategies and within the specified resource bounds. This focused model highlights exactly the critical interactions. Finally, the tool synthesizes and returns this same pair of strategies as the winning coalition profile, namely 
\(
s_{\mathrm{carrier}}^{*} = \{(\neg(\mathrm{dist}<0.5),\rightarrow),(\top,\uparrow)\}
\)
and 
\(
s_{\mathrm{villain}}^{*} = \{(\mathrm{dist}\ge0.5,\downarrow)\}
\)
, which is guaranteed to satisfy \(\coop{\mathit{carrier}}^{\le 2}_{\le 5}\neg(\mathrm{dist}<0.5)\,\until\,\mathit{safe}\).

\section{\uppercase{Conclusion}}
\label{sec:Conclusion}

In this paper, we have addressed the problem of modeling human‑like strategies in MAS by explicitly accounting for bounded rationality, non-zero action costs, and uncertain/noisy perceptions over the time. We have encoded strategies as concise, rule‑based controllers 
with fuzzy semantics for predicates and real‑valued action costs drawn from a non‑refillable budget. This enables verification of safety and performance under realistic constraints, better orchestration of human–AI handoffs, and auditable explanations (e.g., request confirmation if confidence $\le \tau$ and interrupt cost $\leq c$). The approach is especially relevant in search‑and‑rescue robotics, cybersecurity triage, healthcare and mixed traffic, and critical‑infrastructure operations. Technically, we have introduced \HumanATLF, provided its syntax and cost‑aware fuzzy semantics, and established complexity results: model checking is in \(\mathsf{P}\) when both the strategy complexity bound \(k\) and budget \(b\) are fixed, \(\mathsf{NP}\)‑complete in case of a single strategic operator over Boolean objectives, and \(\Delta^P_2\)‑complete when \(k\) and \(b\) can vary; for recall‑based strategies, a bounded unfolding of the game graph yields a PSPACE decision procedure. Our open‑source implementation in VITAMIN and its evaluation on a suite of benchmarks (including an adversarial drone‑rescue scenario) demonstrate that these guarantees translate into practical performance. 

Future research may explore how to refine our understanding of “naturalness” within the Natural Strategic Abilities paradigm by moving beyond mere condition complexity to encompass richer forms of human simplification. Another significant aspect could be enhancing the expressive power of the logic, incorporating elements such as knowledge operators, belief systems, and strategic hierarchies would improve psychological realism. Finally, to overcome the combinatorial explosion of exhaustive strategy generation and scale our approach to larger real‑time models, we will leverage neurosymbolic AI techniques, integrating large language models to optimize strategies generation. These extensions will broaden our framework’s applicability in the already cited resource‑constrained Multi‑Agent domain.

\section{Acknowledgments}
This research is supported by the following projects:  ANR project NOGGINS ANR-24-CE23-440, PRIN 2020 RIPER - CUP E63C22000400001, ECS00000037-MUSA-INFANT, ECS00000009-Tech4You-APLAND, and SHINE-QC Prog n. F/350362/03/X60 - CUP: B69J25000260005 - COR: 24282027. 

\bibliography{aaai2026}

\newpage

\end{document}